\documentclass[
notitlepage,
aps,
prl,
12pt,
superscriptaddress,
amsmath,
amssymb,
]{revtex4-1}

\usepackage{times}
\usepackage[T1]{fontenc}
\usepackage[utf8x]{inputenc}
\usepackage{amsmath}
\usepackage{amssymb}
\usepackage{amsfonts}
\usepackage{amsthm}
\usepackage{float}
\usepackage{bm}
\usepackage{latexsym}
\usepackage{hyperref}
\usepackage{multirow}
\usepackage[capitalize]{cleveref}
\usepackage{graphicx}
\usepackage{float}
\usepackage{geometry}
\hypersetup{
    colorlinks,
    citecolor=blue,    filecolor=blue,
    linkcolor=blue,    urlcolor=blue
}
\usepackage{lineno}
\usepackage{caption}
\begin{document}


\title{Experimental demonstration of mice tumor control with a laser-accelerated high-energy electron radiotherapy prototype}

\author{Zhiyuan Guo\footnote{These authors contribute equally}}

\affiliation{Department of Engineering Physics, Tsinghua University, Beijing 100084, China}

\author{Shuang Liu\footnotemark[1]}

\affiliation{Department of Engineering Physics, Tsinghua University, Beijing 100084, China}

\author{Bing Zhou\footnotemark[1]}

\affiliation{Department of Engineering Physics, Tsinghua University, Beijing 100084, China}
\affiliation{School of Physics, Zhengzhou University, Zhengzhou, 450001, China}

\author{Junqi Liu\footnotemark[1]}
\affiliation{Department of Radiation Oncology, The First Affiliated Hospital of Zhengzhou University, Zhengzhou 450052, China}

\author{Haiyang Wang\footnotemark[1]}
\affiliation{Department of Radiation Oncology, The First Affiliated Hospital of Zhengzhou University, Zhengzhou 450052, China}

\author{Yang Wan}
\email[]{wany12thu@gmail.com}
\affiliation{School of Physics, Zhengzhou University, Zhengzhou, 450001, China}
\affiliation{Beijing Academy of Artificial Intelligence, Beijing 100084, China}

\author{Yifei Pi}
\affiliation{Department of Radiation Oncology, The First Affiliated Hospital of Zhengzhou University, Zhengzhou 450052, China}

\author{Xiaoyan Wang}
\affiliation{Department of Radiation Oncology, The First Affiliated Hospital of Zhengzhou University, Zhengzhou 450052, China}

\author{Yingyi Mo}
\affiliation{Department of Radiation Oncology, The First Affiliated Hospital of Zhengzhou University, Zhengzhou 450052, China}

\author{Bo Guo}
\affiliation{Beijing Academy of Quantum Information Sciences, Beijing 100193, China}

\author{Jianfei Hua}
\affiliation{Department of Engineering Physics, Tsinghua University, Beijing 100084, China}

\author{Wei Lu}
\email[]{weilu@mail.tsinghua.edu.cn}
\affiliation{Department of Engineering Physics, Tsinghua University, Beijing 100084, China}


\begin{abstract}
\textbf{Radiotherapy using very-high-energy electron (VHEE) beams (50-300 MeV) has attracted considerable attention due to its advantageous dose deposition characteristics, enabling deep penetration and the potential for ultra-high dose rate treatment. One promising approach to compactly delivering these high energy electron beams in a cost-effective manner is laser wakefield acceleration (LWFA), which offers ultra-strong accelerating gradients. However, the transition from this concept to a functional machine intended for tumor treatment is still being investigated. Here we present the first self-developed prototype for LWFA-based VHEE radiotherapy, exhibiting high compactness (occupying less than 5 m$^2$) and high operational stability (validated over a period of one month). Subsequently, we employed this device to irradiate a tumor implanted in a mouse model. Following a dose delivery of $5.8\pm0.2$ Gy with precise tumor conformity, all irradiated mice exhibited pronounced control of tumor growth. For comparison, this tumor-control efficacy was similar to that achieved using commercial X-ray radiotherapy equipment operating at equivalent doses. These results demonstrate the potential of a compact laser-driven VHEE system for preclinical studies involving small animal models and its promising prospects for future clinical translation in cancer therapy.}
\end{abstract}

\maketitle

\normalfont

\section{Introduction}

Nowadays, cancer curability remains a significant global challenge despite constant advances in early diagnosis and treatment options. Radiation therapy, applied to more than half of the patients, is an essential part of cancer treatment that relies on the use of ionizing radiation such as photons, electrons, protons and ions to locally deposit dose to damage cancer cells. Currently, the widely applied radiotherapy facilities still mainly rely on high-energy photon beams. With the development of new technology such as Intensity modulated radiotherapy (IMRT)\cite{BENTZEN2005Intensity} or volumetric modulated arc therapy (VMAT)\cite{Otto2007Volumetric}, the doses can be delivered to tumors more precisely than a few decades ago, whereas the risk of exposure of the surrounding normal tissues remains a concern for patient outcomes. Proton therapy\cite{Wilson1946Protons,Smith2006Proton}, on the other hand, has been considered to exhibits more dose conformality due to the Bragg peak effect. However, the sychrotron/cyclotron machines to accelerate proton beams are large and expensive, and the dose distribution is sensitive to tissue density inhomogeneity, resulting in range uncertainties and potential over-irradiation of healthy tissues\cite{Urie1986Degradation,Flatten2019Quantification}.

Over the past two decades, very high-energy electrons (VHEE) beam, ranging in energies from 50 to 300 MeV, has been considered as a promising alternative approach for the treatment of deep-seated tumors\cite{DesRosiers_2000,Ronga2021future}. Compared with photon and proton beams, VHEE beams offer several benefits, including the insensitivity to nonuniform density\cite{Papiez2002therapy,LAGZDA2020heterogeneous}, easy to be manipulated by small magnetic components\cite{Malka_2005_LPA_VHEE,Svendsen2021LPA-VHEE,whitmore2023CERN}, and potentially more compact and significantly cheaper than current installations required for proton therapy. Recently, the rapid development of laser wakefield accelerator (LWFA) has brought new prospects for building compact and cost effective VHEE radiotherapy machines \cite{Malka2008compact,joshi2020perspectives}. In LWFA\cite{LPA_1979}, electrons are accelerated by the plasma wave driven by a short intense laser pulse. Being free of the metallic breakdown, LWFAs can sustain accelerating fields with more than three orders of magnitude stronger than conventional radio-frequency accelerators, resulting in several hundreds MeV energy gain of electrons within only a few millimeters distance \cite{mangles2004monoenergetic,geddes2004high,faure2004laser}. Many research groups worldwide have been dedicated to the LWFA-based VHEE field, and progresses on dosimetry\cite{lundh2012comparison,Svendsen2021LPA-VHEE,labate2020toward}, treatment planning\cite{fuchs2009treatment,Svendsen2021LPA-VHEE} and radiobiological effect\cite{oppelt2015comparison,small2021evaluating} have been made. However, explorations towards clinical translation are still at early stage, nor has a dedicated LWFA-based VHEE machine with high performance and stability required for preclinical studies been established.

In this article, we present the first LWFA-based prototype addressing clinical needs for VHEE therapy in the following aspects. First, occupying less than 5 m$^2$ in size, this machine compactly consists of a self-built industrial-grade 20 TW laser system, electron acceleration module, beam transport module, dose shaping module, and the final platform for dosimetry and small animal treatment, easily adapted to a common photon-therapy room. Second, stable operation for both daily performance over an entire month (22 weekdays in total, 2000-3000 shots per day), and also continuous running for tens of hours has been tested, ensuring its feasibility for clinic applications. Third, a dedicated dose delivery with prescribed homogeneous dose distributions and sub-millmeter pointing fluctuation was achieved. We further applied this prototype to perform a small animal irradiation study. After 5.8$\pm$0.2 Gy was delivered by the LWFA electron beams, significant tumor control was observed for all irradiated mice in the following few weeks, and the efficacy was shown comparable with the commercial photon radiotherapy devices using equivalent doses.

\section{The LWFA-based VHEE therapy prototype}

\begin{figure}
  \centering
  \includegraphics[width=\textwidth]{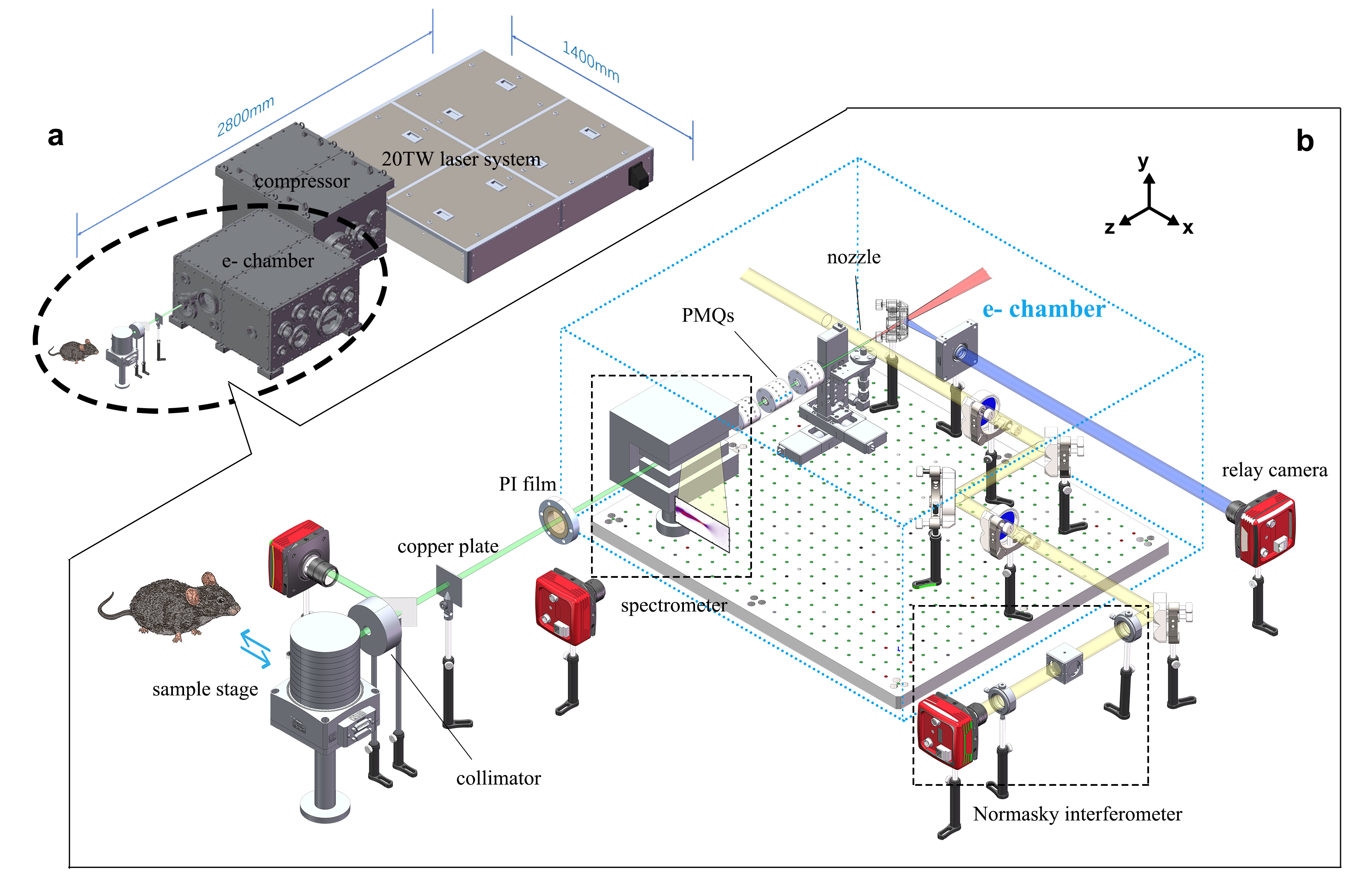}
  \caption{\textbf{The LWFA-Based VHEE therapy prototype.} \textbf{a}, general layout of the device
  , measuring 2.8 meters by 1.4 meters in size.
  \textbf{b}, the detailed setup contained within the dashed circle in \textbf{a}, where the 20 TW laser pulse was focused onto a gas jet to generate a high-energy electron beam with energy up to 175 MeV. The laser focus was assessed using a relay camera, and the plasma density information was examined using a synchronized, low-energy laser pulse directed to a Normasky interferometer. A PMQ triplet was employed to re-collimate the electron beam, reduce pointing variations and eliminate low-energy electrons, which was followed by a permanent dipole magnet to measure the electron spectrum. After the e$^-$ chamber's exit, a scattering plate and a collimating aperture were introduced to shape the beam profile before it reached the sample stage. During the experiment, either animal samples or phantoms were placed on the stage.} \label{fig:graph}
\end{figure}

\textbf{Physical layout.} The general construction of the presented LWFA-based VHEE therapy machine is presented in Figure 1a. To meet the demand for industrial-level LWFA applications, the 20 TW laser system has undergone a decade of independent development by our group to achieve high stability and high compactness. For an 8-hour running with 10 Hz repetition rate, the energy stability can reach less than 0.7\% (RMS) and angular pointing less than 2 $\mathrm{\mu}$rad (RMS). The spatial size reaches only 1.4 m by 1.3 m, much more smaller than other commercial products from companies such as Thales and Amplitude Technology. More details about the laser system can be found in the Methods and Supplementary.

Following the main laser chain, the 800 nm laser pulse was compressed to about 25 fs by the pair of gratings in the compressor chamber and then guided to the e$^-$ chamber for accelerating electron beams. There, it was focused by an off-axis parabolic (OAP) mirror onto a supersonic gas jet with an on-target energy of around 490 mJ. A laser probe beam was directed perpendicular to the main laser pulse, passing through the nozzle, and carrying the plasma density information which was then measured using a Nomarski interferometer. After exiting the gas jet, the electron beam propagated through a beam transport module consisting of three permanent magnet quadrupoles (PMQs) to reduce the pointing jitter. Before the electron beam exited the vacuum chamber, a movable permanent magnet dipole was installed for measuring the beam energy spectrum.

A thin film made of polyimide with a thickness of 100 microns was placed on the exit flange, which served as a barrier between the atmosphere and the vacuum while minimizing its impact on the electron beam. After the electron beam entered the atmosphere, it passed through a dose shaping module consisting of a scattering plate and a collimating aperture to control the dose spatial distribution and then irradiated on the samples. The entire treatment system's dimensions were 2.8 m by 1.4 m.

\begin{figure}
  \centering
  \includegraphics[width=\textwidth]{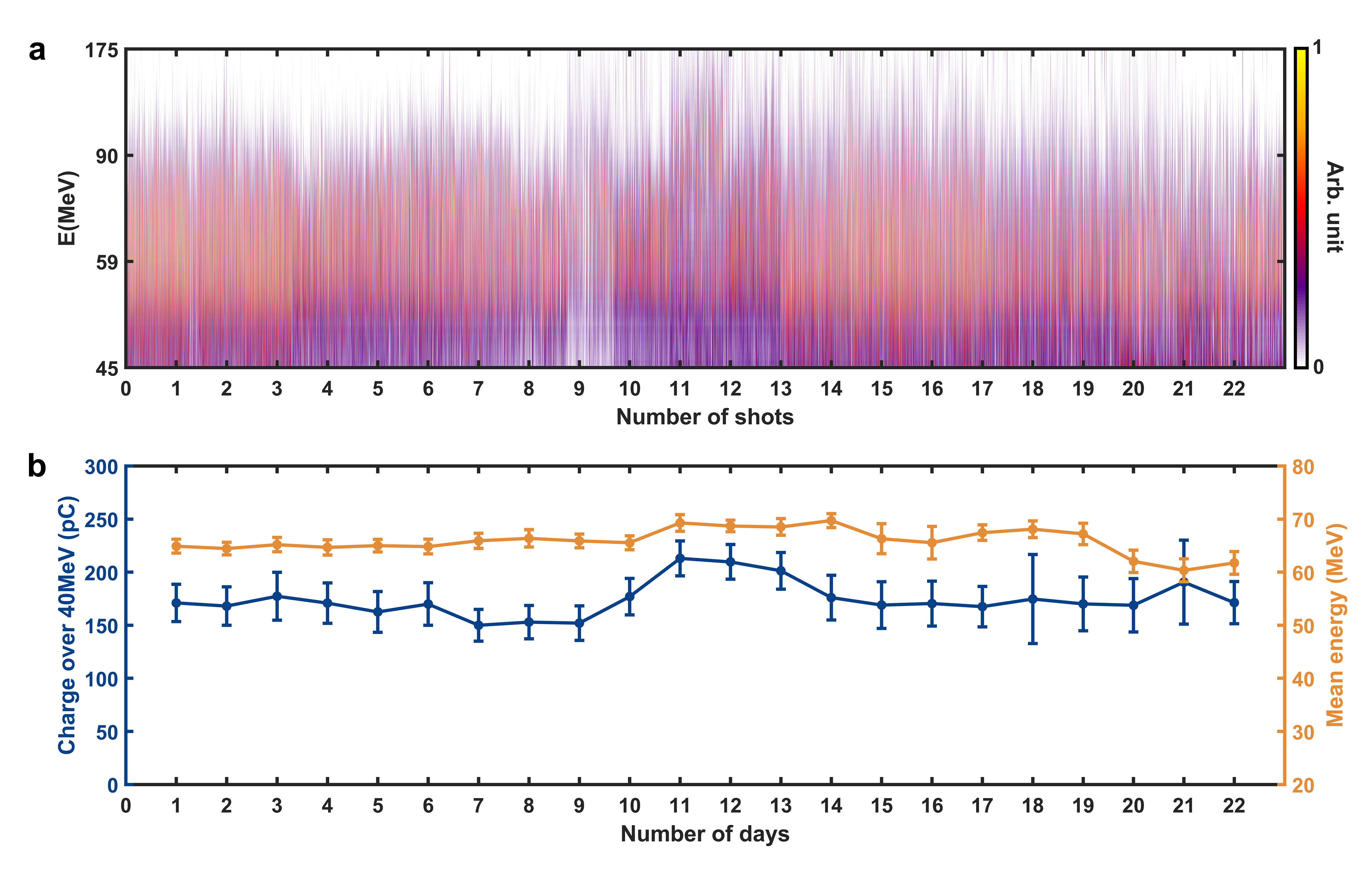}
  \caption{\textbf{Long-term stability of the prototype machine.} \textbf{a}, Sampled electron beam spectra throughout April of the year 2023. The total number of shots recorded is 2522, with 100-130 shots sampled each day. \textbf{b}, Averaged charge (blue) and mean energy (yellow) of the spectrum sampled each day, only the portion of the electron beams with energy higher than 40 MeV were measured.}
  \label{fig:graph}
\end{figure}

\textbf{Operation stability.} By using a gas mixture of helium containing 1\% nitrogen with a plasma density of $1\times10^{19}$ cm$^{-3}$, stable and high-charge (>100 pC) electron beams were obtained through ionization injection\cite{mcguffey2010ionization,pak2010injection}. The beam properties including the energy spectra, beam profile, and dose distribution were routinely checked for an entire month (22 weekdays) with 2000-3000 shots each day and the operation rate was 1 Hz.

Figure 2a presents the daily recorded electron beam energy spectra, with a data collection of 100-130 shots per day. We note that though ionization injection in a laser wakefield accelerator usually results in a continuous beam spectrum, in our prototype machine the low energy electrons (<40 MeV) were largely eliminated by a beam transport system, as will be discussed later. Figure 2b provides a statistical analysis of the charge and mean energy distribution for electrons based on Figure 2a, revealing that the daily charge varied between 160 and 210 pC, with a maximum daily fluctuation (RMS) below 15\%, and the mean energy between 60 and 70 MeV, with a maximum daily fluctuation (RMS) below 6.0\%.

In parallel, another experimental test for 10 hours, 0.3 Hz running was also performed and the preliminary results of obtained electron spectra can be found in Supplementary Figure 3, which shows impressively high robustness of the system.

\begin{figure}
  \centering
  \includegraphics[width=\textwidth]{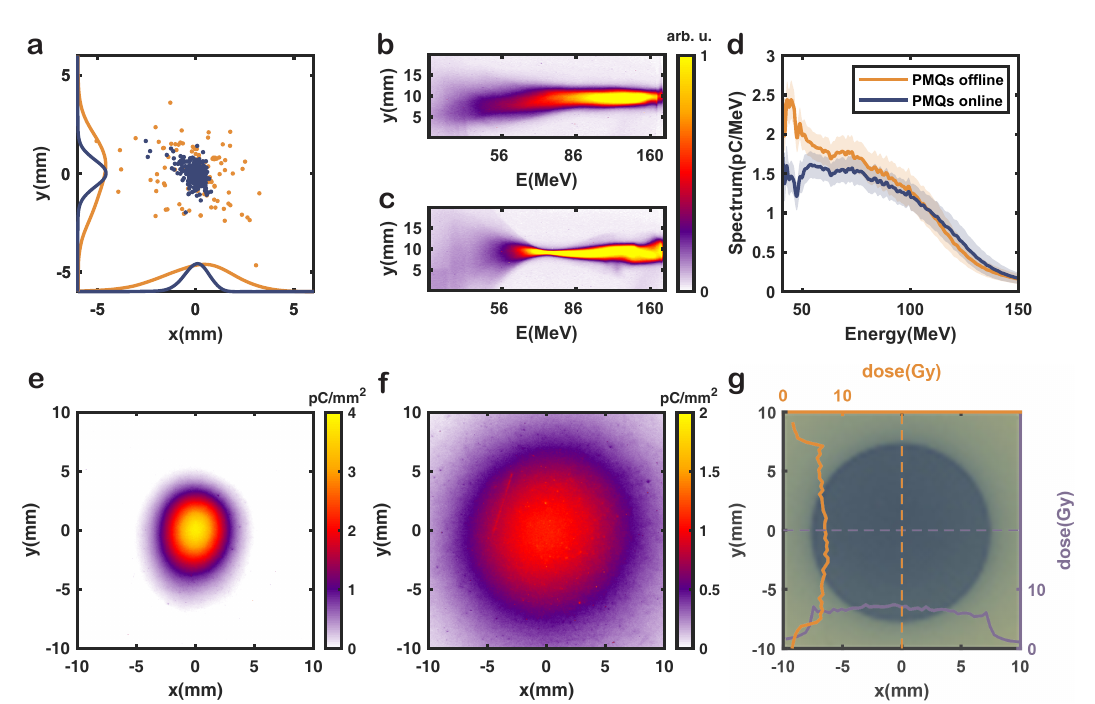}
  \caption{\textbf{Beam transport and dose delivery.} \textbf{a},\textbf{d} Beam pointing jitter and spectrum with (blue) and without (yellow) the transport line. \textbf{b},\textbf{c}, The images on the spectrometer screen without and with the transport line. \textbf{e}, The focused beam profile at the sample position. \textbf{f}, The diffused beam profile at the sample position. \textbf{g}, Dose deposition after the collimator.}
  \label{fig:graph}
\end{figure}

\textbf{Dose delivery.} The pointing jitter of a LWFA electron beam is usually at several mrad level. In our case, as shown in Figure 3a, the standard deviation of the beam center (yellow dots) at the sample position located 1 m away from the plasma source was measured to be 2.4 mm in horizontal ($x$) direction and 2.2 mm in vertical ($y$) direction, which is deficient for accurate dose delivery. To overcome this issue, a beam transport module consisting of three PMQs was employed, which improves the pointing stability of the beam center (blue dots) to 0.31 mm (horizontal) and 0.32 mm (vertical). Furthermore, the beam transport module also exhibits a selective energy effect, reducing the transmission efficiency of the low-energy portion to minimize the radiation dose to the sample/animal skin. Comparing the spectra between Figure 3b and Figure 3c, one can see that the electron beam after passing through the transport module, had a spectrum with a characteristic butterfly-shaped pattern (the focal point was at about 70 MeV) and the low-energy portion was largely divergent. From Figure 3d, one can see electrons with energies above 40 MeV experienced only minimal loss (less than 10\%)  which was cross-verified by numerical simulations. Electrons with energies below 40 MeV, according to the simulations, experienced a loss more than 75\%. More details about the simulations can be found in the Methods and Supplementary.

To study the radiobiological effect, achieving a large and uniform dose distribution covering the entire space occupied by the tumor was essential. However, at the sample position, the laterally focused beam exhibits a Gaussian-like distribution with a FWHM size of approximately 4.1 mm (horizontal) and 4.9 mm (vertical) (see Figure 3e). To achieve a larger lateral irradiation size with uniform dose deposition to cover the 5-7 mm mice tumor, we placed a 1.45 mm copper scatter plate in front of the sample, and the scattered electron beams form a Gaussian-like distribution with a FWHM size of 19 mm at the sample position (see Figure 3f). To further localize the tumor position and minimize the radiation dose to surrounding normal tissues, we adopted a 3 cm-thickness collimating aperture (stainless steel) with a diameter of 15 mm in close proximity to the phantom or biological sample. As shown in Figure 3g, a circular irradiation area with the same size as the aperture and a near-flat top distribution was realized, and the relative dose fluctuation in the flat-topped region was only 11 \%, ensuring the effectiveness of the biological experiments.

\begin{figure}
  \centering
  \includegraphics[width=0.5\textwidth]{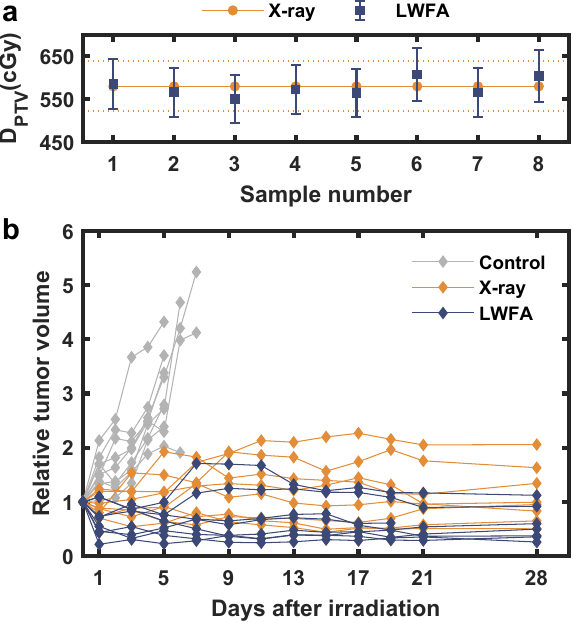}
  \caption{ \textbf{Tumor control in the irradiated mice.} \textbf{a}, The yellow solid line represents the dose received by mice using the medical X-ray radiotherapy machine (Varian Edge series) with the dashed line indicating the uncertainty of the EBT3 film measurements. The blue square represents the dose received by the laser-plasma accelerator system. \textbf{b}, thechanges in tumor volume over a period of 29 days for three groups of mice. The gray line represents the tumor volume changes in the control group mice, the yellow line represents the tumor volume changes in mice treated with a medical X-ray machine, and the blue line represents the tumor volume changes in mice treated with a laser-plasma accelerator.}
  \label{fig:graph}
\end{figure}

\section{Mice in-vivo tumor irradiation}
5-to-6-week-old male C57BL/6JNifdc mice of a total number of 24 were used for the tumor control experiment, which were subcutaneously injected with mouse-derived liver cell carcinoma H22 in their hind legs. These mice were equally divided into three groups: the mice in the control group received regular feeding and were not subjected to any radiation; the mice in the LWFA irradiation group were exposed to radiation from the LWFA-based VHEE. The mice in X-ray irradiation group underwent irradiation using a medical X-ray radiotherapy machine (Varian Edge series). All mice received injections of cancer cells on the same day, and after a period of 7 days, tumors grew to approximately 5-7 mm in size. Mice from both the LWFA irradiation group and the X-ray irradiation group were subjected to radiation on the same day.

In the LWFA-based VHEE irradiation experiments, the mice in the experimental groups were anesthetized and securely fixed onto an anatomical holder. The electron beam was then precisely targeted at the location of the tumor with a nearly uniform dose distribution (see Figure 3g). To induce effective tumor control, the dose administered to each mouse was prescribed to be $5.8\pm0.2$ Gy, measured using EBT3 films placed directly onto the mice. In comparison, the irradiation dose administered to the mice in X-ray group was 5.8 Gy, also measured by a EBT3 film at the surface of each mouse's skin. Due to the characteristic of build-up region in X-ray dose deposition, a 1 cm compensation was placed above the mouse skin to ensure sufficient irradiation to the tumor.

Over a period of 29 days following irradiation, we measured the tumor size in mice and calculated the tumor volume. Figure 4b shows the daily tumor volume evolution normalized by the volume on the irradiation day. The complete data of all the 24 mice in the experiment is summarized in the supplementary material. We observe that despite no significant decrease in the overall body weight of the mice, the tumor size of mice was effectively controlled by irradiation from the LWFA-based VHEE, with similar efficacy compared to commercial X-ray radiotherapy device.

\section{Conclusion and perspective}
In conclusion, we have demonstrated a sustained and stable operation of a LWFA-based VHEE beam on a self-constructed remarkably compact prototype. Employing this prototype, we carried out in vivo experiments targeting live mice tumors and verified that the impact of LWFA-VHEE beam on tumor growth inhibition is prominent and comparable to the effects induced by commericial medical X-ray radiotherapy machines. The presented device, characterized by its occupational compactness,  operational stability, uniform dose delivery, and evident radiobiological effects, emerges as a crucial platform for advancing translational research in the field of LWFA-based radiation oncology.

In the future, we plan to integrate machine learning techniques into our system to enhance both beam quality and stability \cite{Jalas2021Bayesian,Kirchen2021loading}. We also aim to explore innovative diagnostic methods, including femtosecond relativistic electron microscopy \cite{zhang2017femtosecond,wan2022direct,wan2023femtosecond}, to achieve a more comprehensive understanding of the electron acceleration process. Currently, the presented prototype offers electron beams in a fixed direction, and our upcoming developments will involve the implementation of multi-angle treatment approaches, such as a rotating gantry, coupled with novel treatment planning systems tailored for VHEE. This strategic move aims to further minimize radiation doses to healthy tissues while leveraging the inherent advantages of VHEE for treating deep-seated tumors. Moreover, given the ultrafast nature of LWFA electron beams, ongoing exploration into the potential for achieving the ultra-high dose rate effect in LWFA-based VHEE remains a focal point of our research.

\section*{Methods}

\paragraph{\bf Laser system.}

In our prototype device, a 20 TW laser system was adopted. It was fully self-developed from oscillators, pumps to amplifiers and works at a repetition rate of 10 Hz.  Due to the usage of flexure mirror mounts for optical elements and independent cooling system inside each module, the system reaches very high performance, with a pointing stability better than 2 $\mu$rad (RMS) and an energy stability less than 0.7\% (RMS). Both were tested for 8 hours running at 10 Hz. More details about the characterization of the laser system can be found in the Supplementary.

\paragraph{\bf \small Generation, transport and diagnostic of the LWFA-VHEE beams.} 
The 490 mJ, 25 fs laser pulses were focused to 12.3 $\mu$m and 12.6 $\mu$m (FWHM), corresponding to a peak normalized vector potential $a_0$ of 2.5. A supersonic 2-mm nozzle was employed and filled with a gas mixture of 99\% helium and 1\% nitrogen. The working plasma density was approximately $1\times10^{19}{\rm {cm}}^{-3}$. This configuration of laser pulse and gas source could efficiently generate plasma wakefield in bubble regime\cite{lu2006nonlinear,pukhov2002laser} and induce ionization injection\cite{mcguffey2010ionization,pak2010injection}, which resulted in a broad electron energy spectrum up to 175 MeV.

The beam transport system comprising three PMQs was designed to focus the electron beam into parallel beams for successive irradiation. The transport system was located 8.5 cm behind the source, with a distance of 4 cm between the first and second magnets, and a distance of 2.8 cm between the second and third magnets. The detailed parameters of the PMQs can be found in S-Table.2 of the Supplementary. 

For the energy spectrum diagnosis, a dipole magnet had a maximum magnetic field of 1 Tesla and was positioned 9 cm behind the beam transport system. A calibrated scintillation screen was positioned 9 cm behind the dipole magnet to collect the beam information\cite{Wu2012DRZ}, from which the electron beam spectrum as well as the beam charge can be deduced.

\paragraph{\bf \small Numerical simulation.} We performed the beam transport simulation using the code Transwin to calculate the electron beam envelope evolution (see S-Figure 4 in the Supplementary) and to estimate the spectra tailoring effect of the PMQs (see S-Figure 5 of the Supplementary). In the simulation, The electron beam was initialized at the gas jet with a RMS beam size of $1\mathrm{\mu m}$(approximate point source), and a RMS divergence of $\theta_x=5\mathrm{mrad}$ and $\theta_y=6.8\mathrm{mrad}$. A flat-topped spectrum ranging from 0 to 150 MeV was adopted. This system was capable of filtering 75\% electrons with energies below 40 MeV, while achieving a near 100\% transport efficiency for electrons with energies over 50 MeV. 

In additional, an Monte Carlo simulation using the code TOPAS \cite{Perl2012TOPAS} was carried out to simulate the beam size at the sample position for different configurations. From S-Figure 6 of the Supplementary, one can see that, without passing through the beam transport system, the FWHM beam spot sizes were 12 mm (horizontal) and 9 mm (vertical). After being focused by the triplet, they narrowed down to 4 mm (horizontal) and 3 mm (vertical). Subsequently, after being scattered by the scattering plate, the beam distribution became more uniform, with FWHM sizes of 2 cm (horizontal) and 1.8 cm (vertical). These simulation results were in good agreements with the experiment (see Figure 3).

\paragraph{\bf \small Dosimetry of the LWFA-VHEE beams}

In the present study, we employed EBT3 film for measuring radiation dose. The principle of dose measurement of EBT3 film is establishing a relationship between the coloration of the irradiated film and the received dose. Prior to the experiment, a medical X-ray radiotherapy machine (Varian Egde series) and an \textit{Epson 12000XL} scanner were utilized to calibrate the film coloration, and the \textit{Film QA pro} software was utilized for calibrating the dose-response curve and reading the dose results. The EBT3 films were scanned 24 hours after irradiation.

\paragraph{\bf \small Tumor model}

The animal facilities and the experiments were approved according to the regulations on the management of experimental animals in China and the local ethics committee (approval 2023-KY-0703-001, The First Affiliated Hospital of Zhengzhou University, Henan, China). The experiments were performed using 5-to-6-week-old male C57BL/6JNifdc mice purchased from Beijing Vital River Laboratory Animal Technology Co., Ltd. one week prior to the tumor injection. The animals were kept grouped with a maximum of four mice per cage at $12:12 h$ light–dark cycle, constant temperature of about 26 $^{\circ}$C, and relative humidity of 70\%. The mice were fed with a commercial laboratory animal diet and water ad libitum. Studies were carried out for the mouse-derived liver cell carcinoma H22. tumor growth was measured every day using a caliper. The corresponding tumor volumes were calculated as $ \frac{\pi a b^2}{6}$, where a is the longest tumor axis and b is the shortest tumor axis perpendicular to a. tumor-bearing animals that met the eligibility criteria were randomly assigned to different treatment groups in order to distinguish the temporal variations of the tumor model and non-radiation effects from the treatment outcomes.

\paragraph{\bf \small Irradiation using a clinical X-ray radiotherapy machine}

In this experiment, we used high-energy X-rays generated by a medical linac Varian Edge to irradiate mice in order to compare the radiobiological influence between conventional medical radiotherapy and LWFA-VHEE radiotherapy. The X-ray is generated due to the bremsstrahlung effect of 6 MeV electron beams. The medical X-ray linac can generate a uniform dose distribution. In the Linac experimental group, due to the characteristics of X-ray interactions in water, including the build-up effect and backscattering effect, mice were placed between a 1 cm thick tissue compensator and a 5 cm thick water phantom. This setup allowed for more uniform and stable dose delivery to the mice. EBT3 film was also placed between the tissue compensator and the water phantom to monitor the dose received by the mice.

\section*{Acknowledgements}
This work was supported by the National Natural Science Foundation of China (NSFC) Grants (No. 11991071, No. 11775125, No. 11875175, No. 12375241 and No. 12305152), CAS Center for Excellence in Particle Physics, Key Scientific Research Projects of Henan Provincial Colleges and Universities (No. 23B320004), Henan Province Medical Science and Technology Research Program Provincial and Ministry Co-constructed Youth Project (No. SBGJ202103073).

\section*{Author contributions}
Y.W. and W.L. conceived the idea and designed the prototype setup. The 20-TW laser system were developed by W.L. and his laser team at Beijing Academy of Quantum Information Sciences. S.L., B.Z., Z.G., and Y.W. prepared and conducted the experiments with the support from J.L, H.W., Y.P, J.H. and B.G.. J.L., Z.G., Y.W., X.W. and Y.M. contributed to animal handling and care. Z.G., H.W. and Y.P. contributed to the EBT3 film calibration and dosimetry.
B.Z. performed the Transwin and Monte-Carlo simulations. S.L. and B.Z. contributed to laser operation and maintenance. S.L., Z.G, and B.Z. analysed the experimental data. S.L., Z.G. and Y.W. wrote the manuscript. W.L. supervised the project. All the authors reviewed the manuscript
and contributed to discussions.

\section*{Competing interests}
The authors declare no competing interests.

\section*{Data availability}
All source data and code that support the findings of this study are available upon reasonable request from the corresponding authors. 

\newpage
\bibliography{main_v4.bbl}

\end{document}